\documentclass[12pt,preprint]{aastex}

\newcommand{\Ha}{H$\alpha$}
\newcommand{\ca}{Ca\,{\sc ii} 8542 \AA}
\newcommand{\RHESSI}{$\it RHESSI$}

\shorttitle{WHITE-LIGHT FLARE OF 2002 SEPTEMBER 30}
\shortauthors{CHEN \& DING}

\begin{document}

\title{
FOOTPOINT MOTION OF THE CONTINUUM EMISSION
IN THE 2002 SEPTEMBER 30 WHITE-LIGHT FLARE}
\author{Q. R. Chen and M. D. Ding}
\affil {Department of Astronomy, Nanjing University, Nanjing 210093, China}

\begin{abstract}
We present observations of the 2002 September 30 white-light flare, 
in which the optical continuum emission near the \Ha\ line is
enhanced by $\sim$10\%.
The continuum emission exhibits a close temporal and spatial 
coincidence with the hard X-ray (HXR) footpoint observed by {\it RHESSI}.
We find a systematic motion of the 
flare footpoint seen in the continuum emission;
the motion history follows roughly that of the HXR source.
This gives strong evidence the this white-light flare is powered
by heating of nonthermal electrons.
We note that the HXR spectrum in 10--50 keV is quite soft
with $\gamma \approx 7$ and there is no HXR emission above 50 keV.
The magnetic configuration of the flaring region implies magnetic
reconnection taking place at a relatively low altitude during the flare.
Despite a very soft spectrum of the electron beam,
its energy content is still sufficient to produce the heating 
in the lower atmosphere where the continuum emission originates.
This white-light flare highlights the importance of
radiative backwarming to transport the energy below
when direct heating by beam electrons is obviously impossible.
\end{abstract}


\section{INTRODUCTION}
Compared to ordinary solar flares, 
white-light flares (WLFs) manifest themselves
with an enhanced emission in the optical continuum 
of a few or tens of percent, or even more in some extreme cases.
The continuum emission originates in the lower chromosphere and below.
Therefore, research on WLFs can reveal critical clues to
energy transport and heating mechanisms in the solar lower atmosphere
\citep{nei89, din99}.

Most of the WLFs observed so far belong to Type I WLFs 
that are spectrally characterized with a Balmer and Paschen jump
and strong broadened hydrogen Balmer lines \citep{mac86}.
Observations usually demonstrate 
a close temporal correspondence between the continuum and 
the hard X-ray (HXR) emissions in these WLFs 
\citep[e.g.,][]{hud92, fan95, nei93, mat03, met03, hud05}.
This implies that the continuum emission is closely related to 
nonthermal heating of energetic electrons, 
which are assumed to be accelerated 
in the corona and then stream downward to the chromosphere.
However, this process alone, in most cases, cannot efficiently
produce the continuum emission since very few electrons ($E$ $\geq$ 200 keV)
can penetrate to 
the lower chromosphere and below
\citep{din03b}.
Therefore, radiative backwarming is further invoked 
to transfer the enhanced chromospheric radiation
to deeper layers to produce the heating there 
and finally an enhanced continuum emission
\citep{hud72, abo87, mac89, met90a, met90b,gan94,din03b}.
A recent radiative hydrodynamic model of solar flares,
which includes the electron beam heating and radiative transport 
in a consistent way, reproduces the continuum enhancement
comparable to the observed values \citep{all05}.
On the other hand, observations show that 
in some few very energetic events, electrons can be accelerated to 
very high energies, e.g., 300--800 keV \citep{xu04,xu05}.
In such cases, a large amount of nonthermal electrons can penetrate 
deep into the atmosphere, and direct collisional heating may play an 
important role in producing the continuum emission. 

It has been widely accepted that nonthermal electrons,
accelerated by magnetic reconnection in the corona,
stream downward along magnetic fields and produce the HXR emission
in the chromosphere via bremsstrahlung radiation \citep{bro71}.
According to this scenario, the apparent motions of HXR sources reflect 
the successive reconnection process between 
neighboring field lines in the corona \citep[e.g.,][]{qiu02}.
Observations of footpoint motions can thus provide a test or constraint
on theoretical models of solar flares.
There are, up to now, quite some observations revealing the motions of 
HXR sources indicative of successive reconnection under various
magnetic topologies \citep[e.g.,][]{sak00, kru03}. 
However, few examples have shown the motions of 
white-light continuum kernels that are related with the HXR sources.

In this paper, we present observations of a white-light flare  
on 2002 September 30. 
We examine the temporal and spatial 
relationship between the continuum
emission and the HXR emission during the flare
and discuss its energetics.
In particular,
We discover a fairly good correlation between 
the motion history of the white-light kernel 
and the corresponding HXR source.
We explain the observed continuum emission in terms of  
nonthermal heating by electron beams followed by 
the radiative backwarming effect.

\section{OBSERVATIONS AND DATA ANALYSIS} 
According to the Solar-Geophysical Data,
the flare of 2002 September 30 is an M2.1/1B event that
occurred in NOAA Active Region 0134 (N13\degr, E10\degr).
The profiles of {\it GOES} soft X-ray (SXR) fluxes show that the flare
began at $\sim$01:44 UT and peaked at $\sim$01:50 UT.
The hard X-ray (HXR) emission up to 50 keV of this flare was observed by 
the {\it Reuven Ramaty High-Energy Solar Spectroscopic Imager} ({\it RHESSI}), 
which provides unprecedented high resolution imaging and 
spectroscopy capacity for solar flares \citep{lin02}. 
Note that this flare and the M2.6/2B flare on 2002 September 29
\citep{din03a, che05} are two homologous flares in the same active region.

Using the imaging spectrograph of the Solar Tower 
Telescope of Nanjing University \citep{hua95}
and employing a scanning technique, we obtained 
a sequence of two-dimensional spectra of the \Ha\ and \ca\ lines
across the whole flaring region at a time cadence of $\sim$15 s.
There are 120 pixels with a spacing of 0$\farcs$85 along the slit
and 50 steps with a spacing of 2$\arcsec$ along the scanning direction. 
The spectra contain 260 wavelength pixels with a resolution of 
0.05 \AA\ and 0.118 \AA\ for the \Ha\ and \ca\ lines, and
thus span a range of about 13 \AA\ and 30 \AA, respectively.

During the flare, the line widths of the \Ha\ and \ca\ lines
increase gradually and then decrease after the maximum phase; 
the peak values of FWHM are about 4.5 \AA\ and 1.5 \AA, respectively.
We extract a narrow window in the far red wing of the 
\Ha\ line (e.g., $\Delta \lambda = 6$ \AA) and use it 
as a proxy of the nearby continuum.
We check carefully the line profiles and ensure that the line emission
has little influence on the continuum window.
We define the continuum contrast as $(I_f-I_q)/I_q$, 
where $I_f$ is the flare intensity and $I_q$ the
preflare background (taken about one hour before the flare) 
at the same point.
By checking the mean intensity fluctuation outside the flaring region,
we further estimate the measurement error of the continuum contrast
to be below 1\%.

Reduction of ground-based data includes dark field and flat field corrections. 
We coalign the \Ha\ images in line wings with the
{\it Solar and Heliospheric Observatory} MDI continuum images \citep{sch95}
by correlating the sunspot features. 
The accuracy of image coalignment is estimated to be $\sim$2\arcsec. 

\section{RESULTS AND DISCUSSIONS}
\subsection{ENERGETICS OF THE CONTINUUM EMISSION}
We present in Figure 1 a sequence of images of this flare
as seen in the \Ha\ line center ({\it gray scale}), 
the continuum (\Ha\ + 6 \AA, {\it white contours}),
and {\it RHESSI} 12--25 keV HXR ({\it black contours}).
HXR images are reconstructed with the CLEAN algorithm 
from detectors 3--8, which yields an angular resolution of $\sim$7\arcsec.
Note that the flare is located near the solar disk center and
the flaring loops are very compact. Therefore, 
the \Ha\ or HXR emission from the loop top and the footpoints 
may probably be spatially mixed together due to projection effect.
We also select a 171 \AA\ EUV image from 
the {\it Transition Region and Coronal Explorer} 
({\it TRACE}, \citealt[]{han99})
to show the large structure of the flaring region. 

We investigate the origin of the continuum emission by
examining its relationship with the HXR emission during the flare,
which provides important information
about the heating processes in the flaring atmosphere.
As seen in Figure 1,
the flare has a simple compact morphology
in the continuum and the HXR emission,
both of which reside at the main \Ha\ ribbon;
there is a close spatial coincidence between
the continuum and the HXR emission.
We also examine the temporal variation of the continuum contrast
at the position denoted with the $plus$ sign in Figure 1 where
the maximum continuum contrast appears during the flare.
As seen in Figure 2, the continuum contrast reaches a maximum of 
$\sim$10\% at $\sim$01:49:17 UT and correlates well with 
the HXR emission during the flare.\footnote{In the far blue wing 
of the \Ha\ line ($\Delta \lambda = -6$ \AA), 
the continuum develops similarly and 
reaches a maximum contrast of $\sim$8\%. 
In a near-infrared window near the \ca\ line, we detect a much 
lower contrast of the continuum ($\sim$2\%).}
Note that for this flare, the time profiles for the HXR emission below 
50 keV are rather gradual and there is no HXR emission above 50 keV. 
Observations from the solar broad-band {\it Hard X-Ray Spectrometer} 
({\it HXRS}, \citealt{far01}) give similar results. 
The close temporal and spatial relationship between 
the continuum and HXR emission in this flare 
indicates that the continuum emission is strongly related 
to nonthermal energy deposition in the chromosphere. 

Current WLF models have proposed electron beam heating plus
the radiative backwarming effect to account for 
the observed continuum enhancement. 
\citet{din03b} and \citet{che05} calculated the continuum contrast
near the \ca\ and the \Ha\ lines, respectively,
as a function of the energy flux of 
the electron beam impacting a model atmosphere.

For this flare, we can apply a single power-law fitting to the
photon spectrum for the HXR source during the maximum phase,
as seen in Figure 3.
The power-law spectrum extends down to $\sim$10 keV and
has an index of $\gamma=7.4$,
which is much larger than the often observed values, e.g., 3--5. 
In the scenario that the above HXR emission is produced via 
thick-target bremsstrahlung 
\citep{bro71} by nonthermal electrons
with an assumed low-energy cutoff of 20 keV,
the nonthermal electron power is derived to be
$\sim$3 $\times\ 10^{28}$ ergs s$^{-1}$ and
the energy flux ($F_{20}$) is estimated to be
$\sim$2.0 $\times\ 10^{10}$ ergs cm$^{-2}$ s$^{-1}$
(an average value from 01:49:00 to 01:50:00 UT).
According to \citet{che05}, an energy flux of
$\sim$2.0 $\times\ 10^{10}$ ergs cm$^{-2}$ s$^{-1}$ can produce
an increase of the continuum emission near the \Ha\ line
of roughly 10\%, which is consistent with
the observed continuum contrast in this WLF.

We should note that the estimation of the energy content
of nonthermal electrons suffers from great uncertainties,
mostly due to the uncertainty of the low-energy cutoff,
especially when the spectrum is very steep.
\citet{sui05} constrained the low-energy cutoff
to a very narrow range of $24 \pm2$ keV for an M1.2 solar flare
by assuming thermal dominance at low energies
and a smooth evolution of thermal parameters from the rise
to the impulsive phase of the flare.
If we take the low-energy cutoff to be 15 or 25 keV,
the energy flux would rise to 6 times or drop to only 25\%
of the above value, respectively.
The latter case could not meet the requirement of 
sufficient heating to account for the continuum contrast.

The observational facts that the continuum contrast reaches
$\sim$10\% and that the HXR spectrum is very soft imply that
electrons of very high energies are not the necessary condition
for the generation of the continuum emission in WLFs.
Nonthermal electrons of intermediate energies deposit
most of their energy in the upper chromosphere, while
the radiative backwarming effect plays the right role in 
sufficiently heating the lower chromosphere and below.

\subsection{FOOTPOINT MOTION HISTORY}
It has been recognized that the apparent HXR footpoint motion
maps the successive reconnection process during solar flares.
Different magnetic configurations around the flaring regions
can result in different patterns of HXR footpoint motions. 
For example, 
using the {\it Yohkoh} HXT observations, \citet{sak00} 
found that the double HXR footpoints move antiparallel
in 7 out of the 14 flares studied.
Since the launch of {\it RHESSI} in 2002 February, 
some more research has been dedicated for this topic.
For example, \citet{kru03} found 
a systematic motion of one HXR footpoint 
nearly parallel to the magnetic neutral line
in the 2002 July 23 X4.8 solar flare.
\citet{liu04} observed an increasing separation of two HXR footpoints
that is predicted by the magnetic reconnection process
in the well adopted solar flare model \citep{kop76}.

Compared to the large number of solar flares detected 
in HXR, SXR, \Ha, etc.,
detection of WLFs is very sporadic because of the fact
that WLFs represent only a small fraction of solar flares; and
most importantly, the white-light continuum emission is usually
constrained in a small area and limited in a time period during
the impulsive phase.
There are few studies on the source motion as seen in the continuum
in the past decades owing to the rare detection of
WLFs and the low cadence of observations.
The flare under study has a relatively simple morphology in
both the HXR and continuum emission, 
which enables us to trace and study the footpoint motion
in both the two wavelengths during the flare.
Comparison of the footpoint motion history in the HXR and continuum
emission sheds new lights and put constraints on the modeling of WLFs.

We plot in Figure 4 
the centroids of the continuum emission ({\it plus sign})
at the time during which ground-based observations were made,
as well as the centroids of the 12--25 keV HXR emission ({\it diamond sign}),
superposed on the MDI longitudinal magnetogram.
Here the centroid refers to the intensity-weighted mean position of 
those parts of the source that have intensity $\ge$50\% of 
the maximum source intensity.
It is very obvious that the continuum source and the HXR source
move in a similar trend: they first move westward and 
then turn back to the east after the maximum phase (at $\sim$01:50:08 UT).
Meanwhile, both the westward and eastward motions are 
nearly parallel to the magnetic neutral line ({\it dashed line}).
After the turnover, the source moves along a line more apart from
the magnetic neutral line. 

To demonstrate the relationship between the footpoint motion
and the magnetic configuration of the flare,
we further conduct potential field extrapolation of the MDI 
magnetogram using the code of \citet{sak82}, 
which can provide rough information about magnetic connectivity 
around the flaring region. 
As shown in Figure 5. the flare is confined to 
a series of low-lying magnetic loops. 
A possible picture for the flare is as follows:
first, the footpoint moves westward as a result of the 
successive reconnection between the low-lying magnetic loops;
then, the reversal of the footpoint motion may be related to 
triggering of nearby larger loops more distant from the magnetic 
neutral line.

Since the HXR emission in 12--25 keV may contain considerable 
thermal contribution from the hot plasma in the flaring loop, 
the HXR centroid mentioned above may represent the weighted 
center of both the footpoint and loop top sources.
However, we believe that the centroid motion is mainly from
the motion of the footpoint source.
For this sake, we also check the source motion in 25--50 keV, 
at which thick-target bremsstrahlung emission is dominant.
The results show that the motion pattern is similar to 
what is found above during the maximum phase 
when the 25--50 keV HXR images are available.

Although there is a very good spatial correspondence of
the continuum emission to the 12--25 HXR emission as the flare goes on,
there still exists a clear offset of 2$\arcsec$--4$\arcsec$ 
between the centroids 
of the continuum and the HXR source, as shown in Figure 4.
Because of the limit of spatial resolution
and uncertainty of image coalignment, 
we cannot draw a clear conclusion whether
the offset is really physical or not.

On the other hand, some studies of two-ribbon flares have shown
that the nonthermal signal revealed in \Ha\ spectra is much more distinct
at the outer edges \citep[e.g.,][]{can93, li04},
where an electron beam streams down along newly reconnected lines
and bombards the chromosphere.
In our case, both the HXR emission and the continuum emission
are found to be more apparent at the outer edges of the main flare ribbon,
which is consistent with the previous results.

\subsection{MAGNETIC RECONNECTION AT LOW ALTITUDES}
Generally, magnetic reconnection and subsequent energy release 
in solar flares are assumed to occur in the corona.
On the other hand, some active phenomena show evidence of local heating in
the lower atmosphere (at chromospheric levels and below), 
such as the Type II WLFs and Ellerman bombs \citep[e.g.,][]{din99, che01}.
In such cases, magnetic reconnection is assumed to take place in
the lower atmosphere where the plasma is much denser and partially ionized.
There is, however, no theory that can deal with the particle 
acceleration in the lower atmosphere. In such levels, the plasma is
highly collisional, and the energy released through magnetic reconnection
is mostly consumed to ionize the plasma. Therefore, we can postulate
that few particles can be accelerated to very high energies.
If the accelerated electrons still follow
a power-law distribution, the slope may be much steeper as compared
to those cases in which reconnection occurs in the corona.

Now we discuss the details of the flare on 2002 September 30.
First, there are some aspects similar to the Type A flares
or Thermal Hot Flares mentioned by \citet{tan87} and \citet{den88}:
the time evolution of the HXR emission below 50 keV is very gradual 
and there is no visible increase of HXR emission above 50 keV;
the HXR spectrum in 10--50 keV follows a nonthermal power-law
with a very steep slope of $\gamma \approx$ 7.
Such features are possibly linked to a higher density at
the energy release site \citep{den88}.
Second, as seen from Figure 4,
the magnetic configuration and
the footpoint motion imply a scenario of magnetic
reconnection taking place in a series of low-lying magnetic loops
during the flare. 
Thus, we propose that the scenario of magnetic reconnection
relatively low in the atmosphere 
may explain the quite soft HXR spectrum observed in this flare.
However, we also note that this explanation is very speculative.

Since we do not find a HXR footpoint pair from \RHESSI\ images, 
we are unable to directly measure the loop size and altitude.
By assuming semicircular loops symmetrically 
straddling over the magnetic neutral line, 
the loop altitude is roughly estimated to be
the mean distance between the footpoint and the neutral line,
i.e., $\sim$2--5\arcsec\ ($\sim$1500--3500 km) 
above the photosphere, which falls in between
the upper chromosphere and the lower corona.
This fact seems to support our conjecture that this flare
occurs in a relative low site.

As shown in \S3.2, during the maximum phase, the HXR spectrum 
follows a power-law that extends down to $\sim$10 keV
and the derived energy flux of beam electrons is sufficient 
to account for the continuum contrast. 
On the other hand, since the HXR spectrum is very soft,
thermal plasma may contribute largely 
to the HXR emission in the low energies, e.g., below $\sim$20 keV.
We show in Figure 2 that the plasma temperature,
which is derived from the  background-subtracted
{\it GOES} 0.5--4 \AA\ and 1--8 \AA\ SXR fluxes
\citep[e.g.,][]{tho85, gar94},
well matches the development of the continuum emission.
\citet{mat03} also showed that some WLFs exhibit a stronger
correlation with the SXR emission than with the HXR emission.
Therefore, another possible origin of the continuum emission, 
i.e., thermal heating plus preferentially the backwarming effect,
is worth investigating in the future.

\section{CONCLUSIONS}
After a synthesising analysis of the WLF on 2002 September 30,
we find that there is a fairly close relationship in both time and space
between the continuum and the HXR emission during the flare.
We discover a footpoint motion seen in the optical continuum;
the motion history follows roughly that of the HXR source.
This gives strong evidence that this WLF is powered by energetic 
electrons followed by the radiative backwarming effect.
The magnetic configuration of the flaring region implies magnetic
reconnection taking place at a relatively low layer 
during the flare. 
The HXR spectrum is thus very soft, possibly, due to a low efficiency of 
electron acceleration at low altitudes. 
However, the energy content of the electron beam is still
enough to produce the heating in the lower atmosphere where 
the continuum emission originates.
Radiative backwarming is of course an important mechanism
to transport the energy below when direct heating 
is obviously impossible.

\acknowledgments
We would like to thank H. Hudson and the referee for 
constructive comments that help improve the paper.
We are grateful to the {\it RHESSI}, {\it SOHO}/MDI, {\it TRACE},
and {\it HXRS} teams for providing the observational data
and to T. Sakurai for providing the
current-free magnetic extrapolation program.
We thank S. Krucker for help in {\it RHESSI} data analysis
and F. F\'arn\'ik for interpreting the {\it HXRS} data.
{\it SOHO} is a project of international cooperation between
ESA and NASA.
This work was supported by NKBRSF under grant G20000784,
NSFC under grants 10025315, 10221001, and 10333040,
and FANEDD under grant 200226.

\clearpage
\begin{figure}
\epsscale{1.0}
\plotone{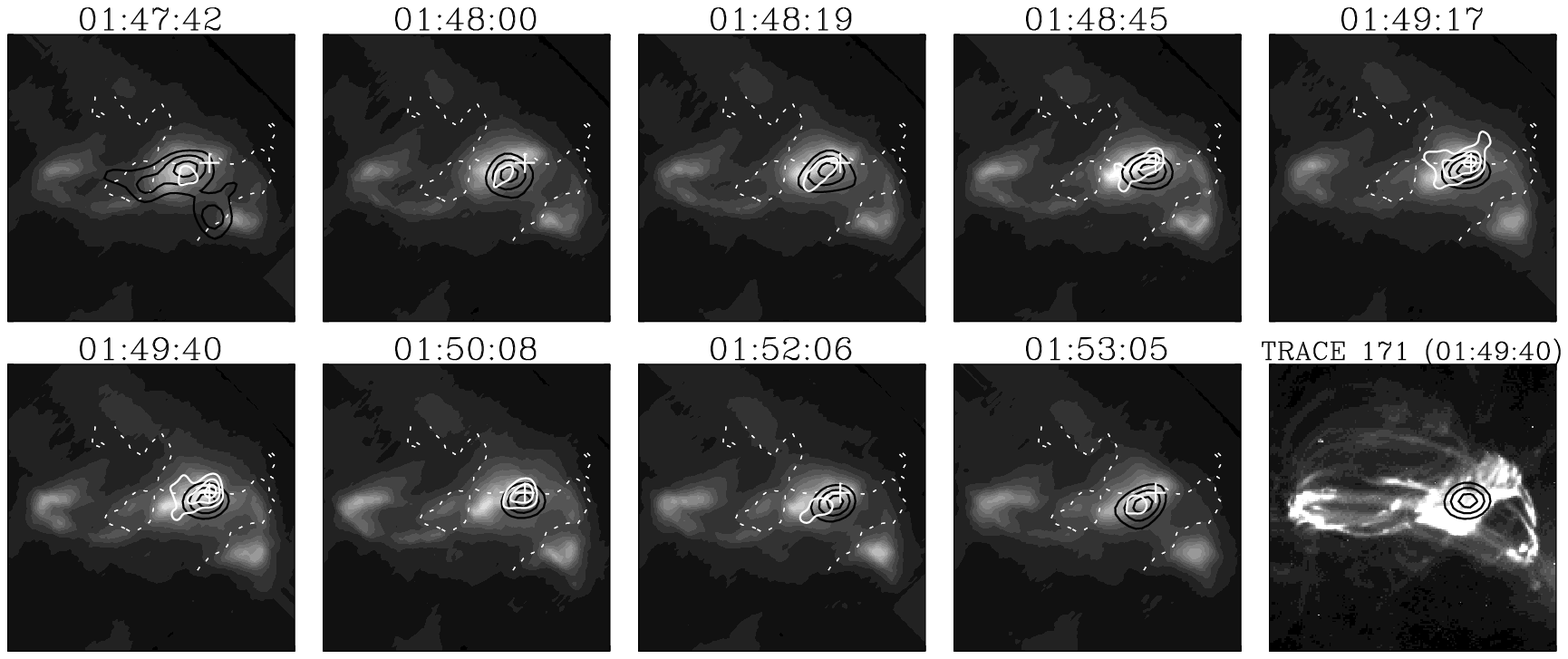}
\caption{
Evolution of the flare in \Ha\ ({\it gray scale}), continuum 
({\it white contours}, with levels of 0.03, 0.06, and 0.09),
and {\it RHESSI} 12--25 keV HXR emission ({\it black contours},
with levels of 50\%, 70\%, and 90\% of the maximum source intensity
in each image)
from 01:47:42 UT to 01:53:05 UT.
The {\it TRACE} 171 \AA\ emission at 01:49:40 UT
is also shown at the lower right panel for comparison.
The white dotted line is the magnetic neutral line of the MDI
magnetogram at $\sim$01:36:01 UT with a correction of 14 minutes
for solar rotation.
The field of view is 90$\arcsec$ $\times$ 90$\arcsec$.
North is up and east is to the left.
\label{fig1}}
\end{figure}

\clearpage

\begin{figure}
\epsscale{0.8}
\plotone{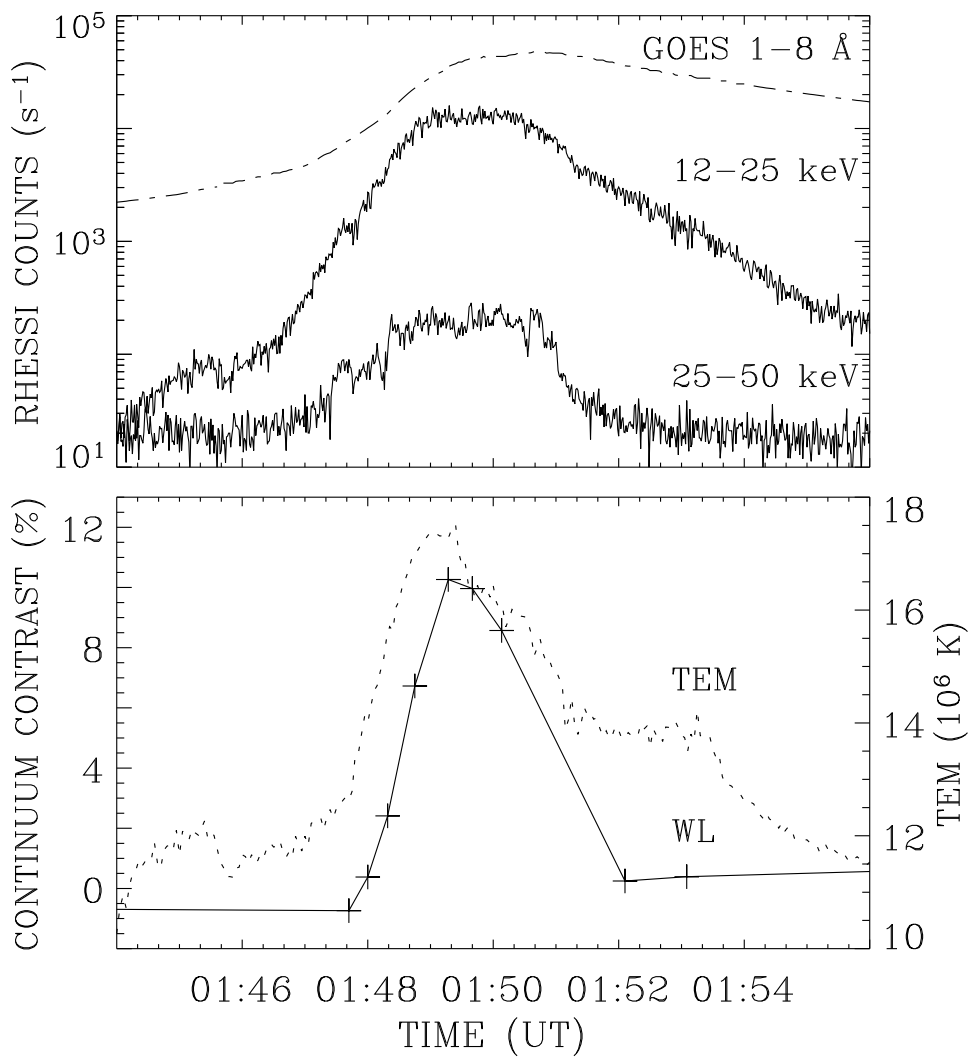}
\caption{
{\it Top panel}:
Time profiles of the {\it RHESSI} HXR emission 
in 12--25 keV and 25--50 keV.
Also plotted is the {\it GOES} 1--8 \AA\ SXR flux 
in arbitrary units ({\it dot-dashed line}).
{\it Bottom Panel}:
Time variation of the continuum contrast 
at the position denoted with the {\it plus} (+) sign in Figure 1.
The {\it dotted line} shows the evolution of the temperature 
derived from the background-subtracted {\it GOES} 0.5--4 \AA\ 
and 1--8 \AA\ SXR fluxes.
\label{fig2}}
\end{figure}
\clearpage

\begin{figure}
\epsscale{1.0}
\plotone{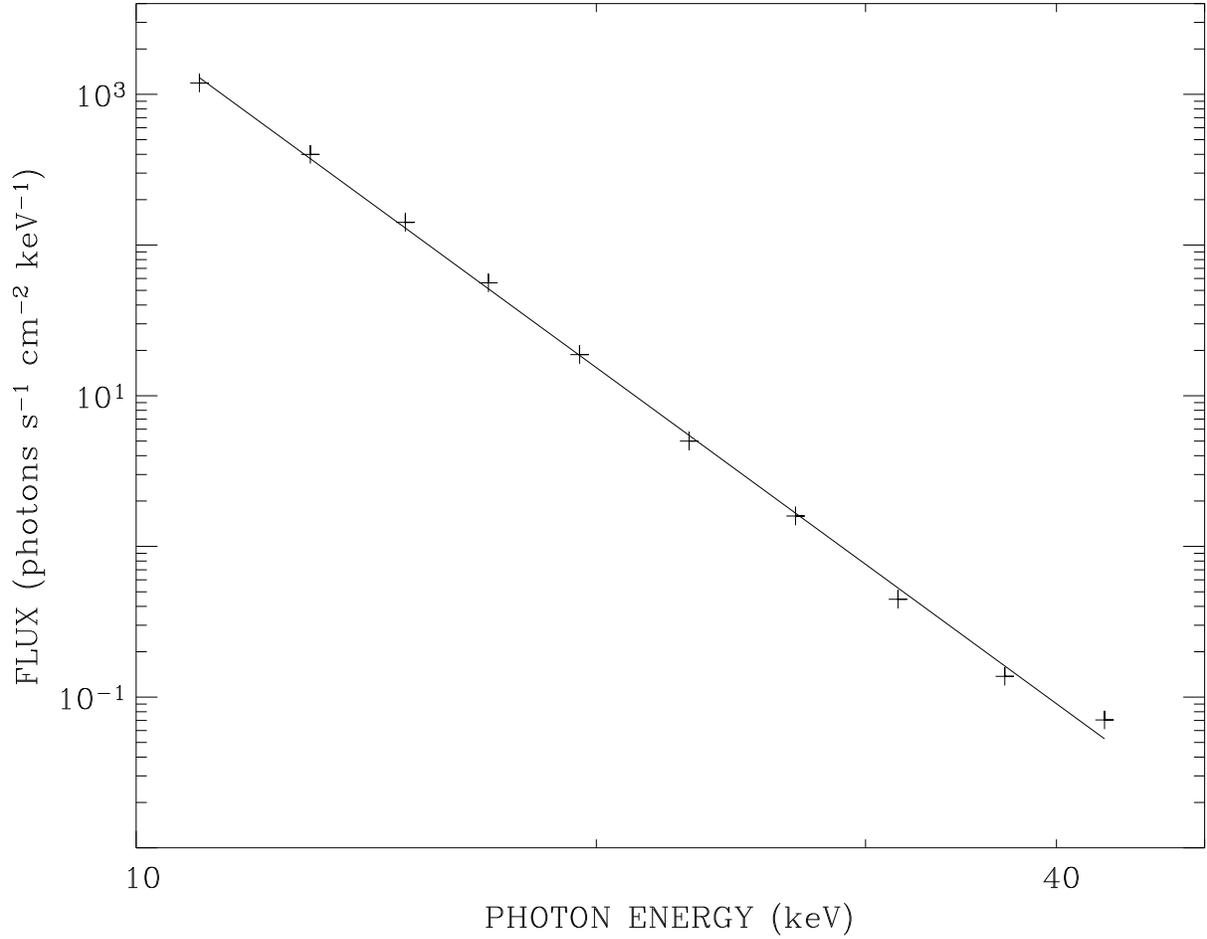}
\caption{
Photon spectrum and its power-law fitting for the
HXR footpoint with an integration time from 01:49:00 to 01:50:00 UT.
The power-law spectrum has a slope of --7.4 and and a flux at 50 keV of 
$\sim$1.73 10$^{-2}$ photons s$^{-1}$ cm$^{-2}$ keV$^{-1}$.
\label{fig3}}
\end{figure}
\clearpage

\begin{figure}
\epsscale{1.0}
\plotone{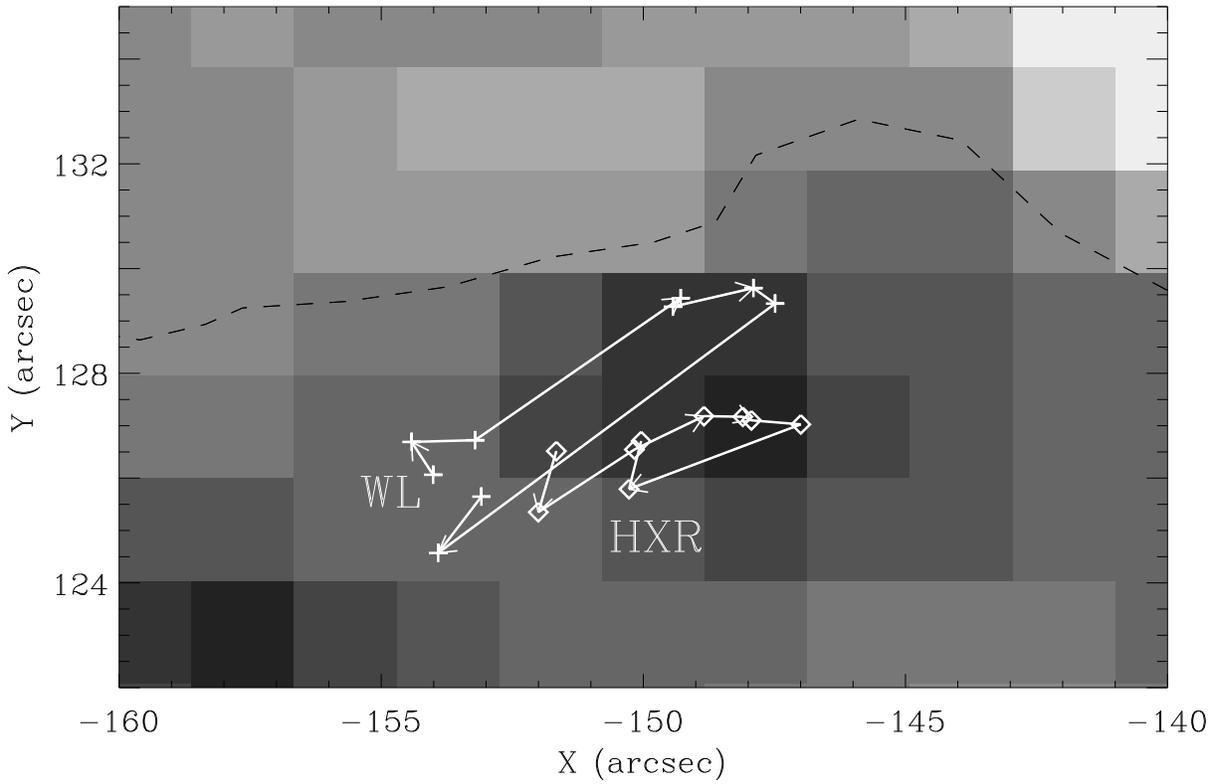}
\caption{
Footpoint motion history in the continuum ({\it plus sign}) and 
12--25 HXR emission ({\it diamond sign}) superposed on the MDI magnetogram. 
The nine time points refer to those times denoted in Figure 1.
The footpoint first move westward and then turns back to the east.
The magnetic neutral line is plotted as {\it black dashed line}.
\label{fig4}}
\end{figure}
\clearpage

\begin{figure}
\epsscale{1.0}
\plotone{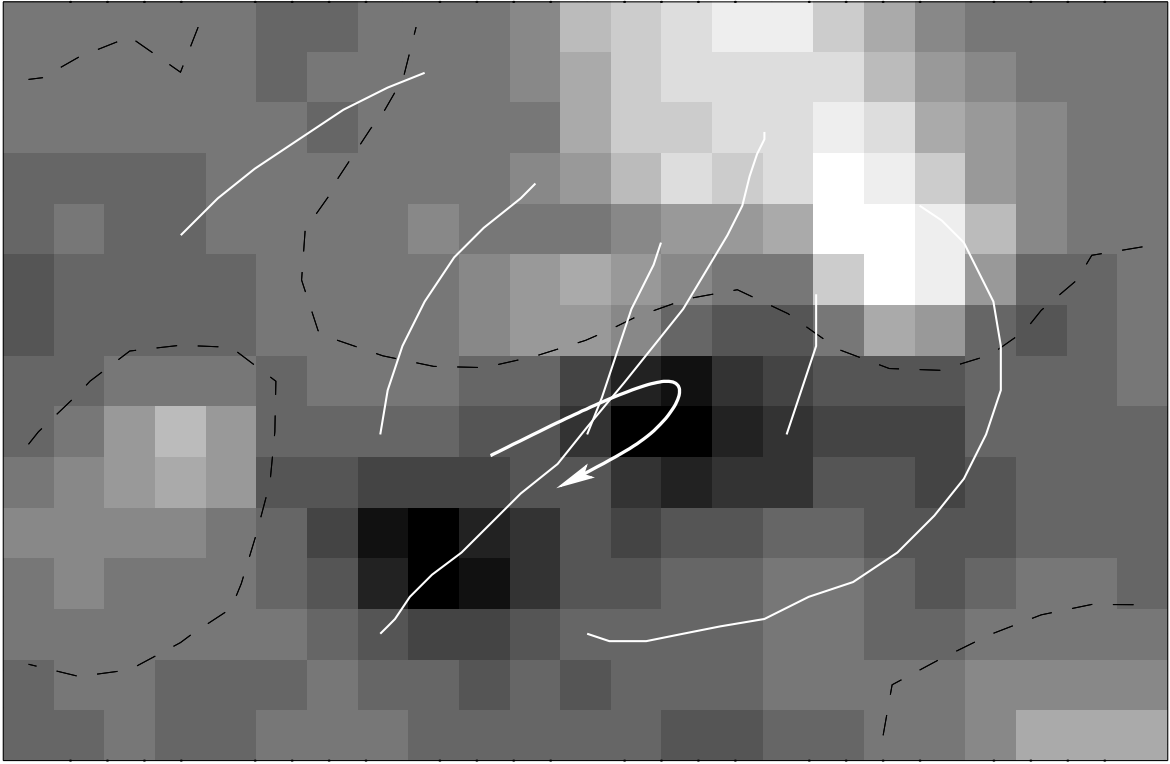}
\caption{
Magnetic field lines ({\it white solid lines}) extrapolated from 
the MDI longitudinal magnetogram ({\it gray image}). 
The footpoint motion trajectory, denoted as a {\it thick white curve},
is superposed on the magnetogram to show its relationship with
the magnetic configurations around the flare.
The field of view is 45$\arcsec$ $\times$ 30$\arcsec$.
North is up and east is to the left.
\label{fig5}}
\end{figure}
\clearpage

\end{document}